\documentclass[conference]{IEEEtran}
\IEEEoverridecommandlockouts
\usepackage{cite}
\usepackage{amsmath,amssymb,amsfonts}
\usepackage{algorithmic}
\usepackage{graphicx}
\usepackage{textcomp}
\usepackage{xcolor}
\def\BibTeX{{\rm B\kern-.05em{\sc i\kern-.025em b}\kern-.08em
    T\kern-.1667em\lower.7ex\hbox{E}\kern-.125emX}}
\begin{document}

\title{Comparative Analysis of Control Strategies for Position Regulation in DC Servo Motors}

\author{
    \IEEEauthorblockN{Raihan Khan Akash} 
    \IEEEauthorblockA{\textit{Electrical and Computer Engineering} \\
    \textit{The University of Alabama in Huntsville}\\
    Huntsville, Alabama, United States \\
    ra0919@uah.edu}
}

\maketitle
\begin{abstract}
A servomotor is a closed-loop system designed for precise movement control, utilizing position feedback to achieve accurate final positions. Due to the ability to deliver higher power output and operate at enhanced speeds, DC servo motors are considered ideal for applications requiring precision and performance. This research aims to design, simulate, and compare various control strategies for precise position control in DC servo motors (DSM). The controllers evaluated in this study include proportional (P), proportional-integral (PI), proportional-integral-derivative (PID), state-feedback controllers (SFC), and state-feedback controllers augmented with integral action (SFCIA). The performance of these controllers was evaluated using MATLAB simulations, characterized by overshoot, settling time, steady-state error, rise time, and peak time. The results indicate that the state-feedback controller with integral action (SFCIA) surpasses other control strategies by achieving zero steady-state error, minimal overshoot, the shortest settling time, and optimized rise and peak times. These findings highlight the effectiveness of SFCIA for tasks requiring high levels of stability, precision, and dynamic performance.
\end{abstract}


\section{Introduction}  
Electric motors, whether AC or DC, are essential components in modern life wherever motion is needed \cite{ref1}. Among them, servo motors particularly stand out for their numerous applications, contributing to the deployment of over 700 million motors worldwide \cite{ref4,ref5}. A servo motor, generally a low-power DC motor acting as an actuator, is designed for precise position control with a range often limited to 180 degrees \cite{ref3}. Despite its compact size, it provides high power, excellent energy efficiency, and impressive dynamic performance due to its low rotor inertia and high torque-to-inertia ratio \cite{ref2}. This combination of reliability, versatility, and cost-effectiveness has made servo motors crucial in the modern world.

Given the vital role in today's life, effective control systems are a must for managing servo motors. Not only do they enhance performance and efficiency but also optimize key dynamic characteristics\cite{ref10}. Over the decades, advancements in control strategies have emerged to meet the growing demand for higher precision, stability, and adaptability. Classical control methods, such as proportional-integral-derivative (PID) controllers, remain popular due to their simplicity but often face limitations like overshoot and poor adaptability under changing conditions \cite{ref7,ref8}. Modern alternatives, including fuzzy logic and state-feedback controllers, offer improved performance but introduce complexities in tuning and implementation, particularly in real-time systems \cite{ref9,ref15}.

The growing interest in overcoming these difficulties has made servo motor control a prominent area of study in recent years. For instance, Dipraj \textit{et al.} (2021) investigated the effectiveness of traditional PI and fuzzy logic controllers for DSM position control using MATLAB/Simulink simulations. Their findings highlighted the limitations of PI controllers in handling overshoot and steady-state error, whereas fuzzy logic controllers showed improved adaptability and performance \cite{ref12}. Pillai and Nair (2017) developed both PID controllers and Model Reference Fuzzy Adaptive Controllers in LabVIEW. They concluded that fuzzy controllers outperformed PID controllers in terms of overshoot and precision \cite{ref13}. Szczepanski \textit{et al.} (2020) introduced an Artificial Bee Colony optimization method to fine-tune state-feedback controllers for two-mass systems, demonstrating significant improvements in system performance \cite{ref14}. Similarly, Mezher (2021) and Maarif and Setiawan (2021) explored advanced PID control and integrated state-feedback approaches, respectively, using MATLAB/Simulink and hardware platforms like Arduino for practical implementation \cite{ref16}.  

Despite these advancements, several challenges persist. Classical controllers such as PID struggle with overshoot, slow response time, and limited adaptability under varying conditions \cite{ref7,ref8}. While fuzzy logic controllers offer flexibility, they are difficult to tune and computationally intensive for real-time applications \cite{ref9}. State-feedback control methods provide precise pole placement capabilities but require careful tuning to ensure system stability, steady-state accuracy, and optimal rise time and peak time \cite{ref15}.  

This study designs, simulates, and evaluates five control strategies: proportional (P), proportional-integral (PI), proportional-integral-derivative (PID), state-feedback control (SFC), and state-feedback control with integral action (SFCIA). MATLAB simulations assess their performance based on metrics such as overshoot, settling time, steady-state error, rise time, and peak time to determine the most effective approach for precise position regulation of DC servo motors under dynamic conditions.

\section{Methodology \& Mathematical Modeling}

\subsection{System Description}
A DC Servo Motor (DSM) is used to control angular position with precision \cite{ref11}. Electrical and mechanical relationships drive the motor's dynamics, and their integration into the mathematical model ensures a complete understanding of the DSM's functionality.

\begin{figure}[h!]
    \centering
    \includegraphics[width=0.5\textwidth]{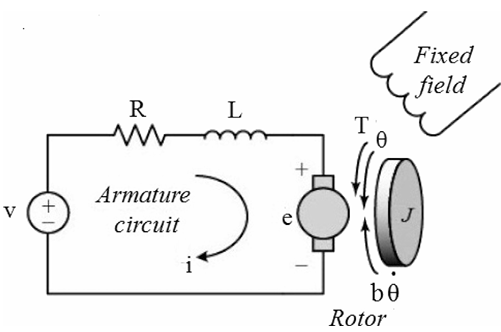}
    \caption{Schematic diagram of the DC Servo Motor showing the armature circuit, fixed field, and rotor dynamics.}
    \label{fig:dc_servo_motor}
\end{figure}

The schematic diagram of the DC servo motor, as shown in Fig.~\ref{fig:dc_servo_motor}, illustrates the armature circuit, fixed field, and rotor dynamics. The armature circuit is represented by a resistor \(R\), an inductor \(L\), and a back emf voltage proportional to the motor speed. The rotor dynamics involve torque generation based on the current and mechanical components like inertia and damping. These relationships are fundamental to formulating the motor's transfer function and state-space representation.
  
\subsection{Electrical Dynamics}
The armature circuit of the DSM is governed by Kirchhoff's Voltage Law (KVL) \cite{ref1}.
\begin{equation}
    V = L \frac{di}{dt} + Ri + e,
    \label{eq:kvl}
\end{equation}
where \( V \) is the armature voltage (input), \( i \) is the armature current, \( L \) represents the armature inductance, \( R \) denotes the armature resistance, and \( e \) refers to the back electromotive force (emf), which is proportional to the angular velocity.

The back emf is given by the equation below.
\begin{equation}
    e = K_e \dot{\theta},
    \label{eq:back_emf}
\end{equation}
where \( K_e \) represents the back emf constant.

\subsection{Mechanical Dynamics}
The torque generated by the motor is proportional to the armature current and is expressed as follows.
\begin{equation}
    T = K_t i,
    \label{eq:torque_current}
\end{equation}
where \( K_t \) is the torque constant. The relationship between torque and the rotor's mechanical dynamics is given by the following euation.
\begin{equation}
    J \ddot{\theta} + b \dot{\theta} = K_t i,
    \label{eq:mechanical}
\end{equation}
where \( J \) represents the moment of inertia, \( b \) is the damping coefficient, \( \ddot{\theta} \) denotes the angular acceleration, and \( \dot{\theta} \) refers to the angular velocity.

\subsection{Transfer Function}
By combining the electrical (\ref{eq:kvl}) and mechanical (\ref{eq:mechanical}) equations, substituting \( i = T / K_t \), and simplifying, the transfer function that relates the input voltage \( V(s) \) to the angular position \( \theta(s) \) is obtained as:
\begin{equation}
    \frac{\theta(s)}{V(s)} = \frac{K}{s \left[ (Ls + R)(Js + b) + K^2 \right]},
    \label{eq:transfer_function}
\end{equation}
where \( K \) is the combined motor constant (\( K_t = K_e = K \)), \( K_t \) is the torque constant, and \( K_e \) is the back emf constant.

\begin{table}[h!]
\centering
\caption{Parameters of the DC Servo Motor}
\begin{tabular}{|l|c|}
\hline
\textbf{System Specifications} & \textbf{Data} \\ \hline
Armature resistance (\( R \)) & 2.45 \( \Omega \) \\ \hline
Armature inductance (\( L \)) & 0.035 H \\ \hline
Torque constant (\( K \)) & 1.2 N-m/A \\ \hline
Back emf constant (\( K \)) & 1.2 V/(rad/sec) \\ \hline
Moment of inertia (\( J \)) & 0.022 Kg-m\(^2\)/rad \\ \hline
Damping coefficient (\( b \)) & 0.0005 N-m/(rad/sec) \\ \hline
\end{tabular}
\label{table:DSM_parameters}
\end{table}

By substituting the DSM parameters provided in Table~\ref{table:DSM_parameters}, the transfer function is derived as:
\begin{equation}
    \frac{\theta(s)}{V(s)} = \frac{1.2}{0.00077s^3 + 0.0539s^2 + 1.441s}.
    \label{eq:theta_transfer}
\end{equation}

\subsection{State-Space Representation}
To facilitate controller design, the dynamic behavior of the DC servo motor is represented in state-space form. This approach provides a compact and flexible representation of the system's dynamics. The general state-space equations are given below.
\begin{equation}
    \dot{x} = A x + B u, \quad y = C x,
    \label{eq:state_space}
\end{equation}
where \( x \) is the state vector, \( u \) represents the input voltage, and \( y \) is the system output.

The state vector and input/output relationships are defined as following.
\begin{equation*}
    x = 
    \begin{bmatrix} 
        \theta \\ 
        \dot{\theta} \\ 
        i 
    \end{bmatrix}, \quad
    u = V, \quad
    y = \text{output}.
\end{equation*}

The system matrices \( A \), \( B \), and \( C \), which depend on the motor's physical parameters, are expressed as:
\begin{equation}
    A = \begin{bmatrix}
        0 & 1 & 0 \\
        0 & -\frac{b}{J} & \frac{K}{J} \\
        0 & -\frac{K}{L} & -\frac{R}{L}
    \end{bmatrix}, \quad
    B = \begin{bmatrix}
        0 \\ 
        0 \\ 
        \frac{1}{L}
    \end{bmatrix}, \quad
    C = \begin{bmatrix}
        1 & 0 & 0
    \end{bmatrix}.
\end{equation}

\subsection{Control Strategies}
In this study, the DSM has been analyzed using various controllers, including Proportional (P), Proportional-Integral (PI), Proportional-Integral-Derivative (PID), State Feedback Controller (SFC), and State Feedback Controller with Integral Action (SFCIA). The corresponding mathematical formulations and associated matrices are detailed in this section.

\subsubsection{Proportional Controller (P)}
Proportional controllers are the simplest form of feedback control systems, where the output is directly related to the error signal, as described below.
\begin{equation}
    c(t) = K_c e(t),
\end{equation}
where \( c(t) \) is the controller output, \( K_c \) is the controller gain, and \( e(t) \) is the error signal.

\subsubsection{Proportional-Integral Controller (PI)}
Proportional-Integral (PI) controllers incorporate integral action to eliminate steady-state error. The control law is given by the following equation.
\begin{equation}
    c(t) = K_c \left( e(t) + \frac{1}{T_i} \int e(t) dt \right),
\end{equation}
where \( T_i \) represents the integral time constant.

\subsubsection{Proportional-Integral-Derivative Controller(PID)}

PID controllers enhance system performance by introducing derivative action in addition to proportional-integral control. The PID control law is described as:
\begin{equation}
    c(t) = K_c \left( e(t) + \frac{1}{T_i} \int e(t) dt + T_d \frac{de(t)}{dt} \right),
\end{equation}
where \( T_d \) is the derivative time constant.
\begin{table}[h!]
\centering
\caption{Control rule settings using Ziegler-Nichols method.}
\begin{tabular}{|l|c|c|c|}
\hline
\textbf{Controller} & \( K_p \) & \( T_i \) & \( T_d \) \\ \hline
Proportional Controller  & 0.5 \( K_{cr} \) & \( \infty \) & 0 \\ \hline
Proportional Integral Controller & 0.45 \( K_{cr} \) & \( \frac{P_{cr}}{1.2} \) & 0 \\ \hline
Proportional-Integral-Derivative Controller & 0.6 \( K_{cr} \) & \( \frac{P_{cr}}{2} \) & \( \frac{P_{cr}}{8} \) \\ \hline
\end{tabular}
\label{table:control_rule_setting}
\end{table}

The Ziegler-Nichols (ZN) tuning method is used to calculate the proportional gain (\( K_p \)), integral time (\( T_i \)), and derivative time (\( T_d \)) for P, PI, and PID controllers \cite{ref17}. Table~\ref{table:control_rule_setting} provides a summary of the tuning procedure, utilizing \( K_{cr} \) (critical gain) and \( P_{cr} \) (oscillation period) as key parameters.

\subsubsection{State Feedback Controller (SFC)}
\begin{figure}[h!]
    \centering
    \includegraphics[width=0.8\linewidth]{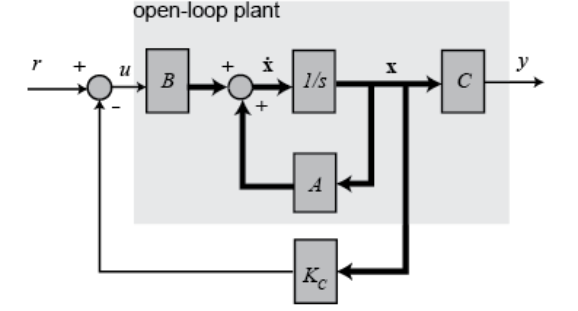}
    \caption{Block Diagram Representing the State Feedback Controller (SFC).}
    \label{fig:SFC_bd}
\end{figure}

The State Feedback Controller (SFC) adjusts the system dynamics by strategically placing the system poles in desired locations. Fig.~\ref{fig:SFC_bd} illustrates the block diagram of the state feedback controller, highlighting how the feedback gain matrix modifies the system input to achieve the desired behavior

The state-space equations for the system with SFC are expressed as follows.
\begin{equation}
    \dot{x} = (A - BK_c)x + Br,
\end{equation}
\begin{equation}
    y = Cx,
\end{equation}
where \( K_c \) is the state feedback gain matrix. The control input \( u \) is defined as below.
\begin{equation}
    u = r - K_cx,
\end{equation}
enabling precise feedback for achieving the desired dynamics.

In this system, all state variables—position (\( \theta \)), speed (\( \dot{\theta} \)), and current (\( i \)) can be directly measured using sensors such as potentiometers, tachometers, and ammeters. This allows for a full-state feedback controller design without the complexity of incorporating an observer, ensuring accurate state feedback.

To meet the desired dynamic performance, closed-loop poles are carefully chosen, and MATLAB's `place` command is used to compute \( K_c \). State feedback provides precise pole placement, enhancing system stability and dynamic response.

\subsubsection{State Feedback Controller with Integral Action (SFCIA)}

\begin{figure}[h!]
    \centering
    \includegraphics[width=0.8\linewidth]{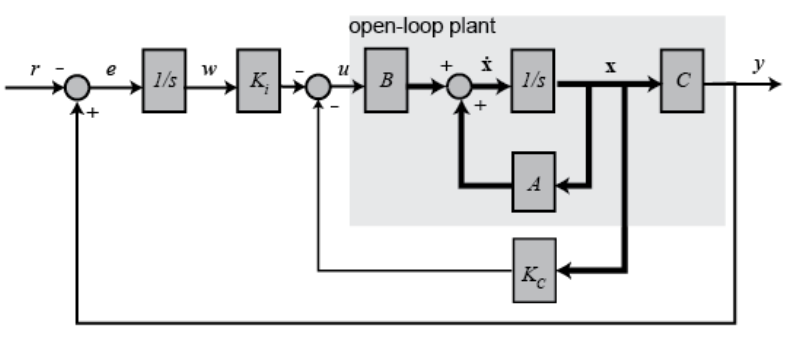}
    \caption{Block Diagram Representing the State Feedback Controller with Integral Action (SFCIA).}
    \label{fig:SFCIA_bd}
\end{figure}

To eliminate steady-state error, integral action is incorporated into the system. The state-space representation is augmented by introducing an additional state variable \( w \), representing the integral of the error, as illustrated in Fig.~\ref{fig:SFCIA_bd}. This modification ensures zero steady-state error while maintaining stability and dynamic performance.

The integral state is defined as follows.
\begin{equation}
    w = \int (y - r) \, dt,
\end{equation}
where \( r \) is the reference input, \( y \) is the system output, and \( e = r - y \) is the error signal.

The augmented state-space equations for the system with integral action can be defined as follows.
\begin{equation}
    \dot{x}_a = A_a x_a + B_a u + B_r r,
\end{equation}
\begin{equation}
    y = C_a x_a,
\end{equation}
where the augmented matrices are expressed as below.
\[
    A_a = \begin{bmatrix}
        0 & 1 & 0 & 0 \\
        0 & -\frac{b}{J} & \frac{K_t}{J} & 0 \\
        0 & -\frac{K_e}{L} & -\frac{R}{L} & 0 \\
        1 & 0 & 0 & 0
    \end{bmatrix},
    \quad
    B_a = \begin{bmatrix}
        0 \\ 0 \\ \frac{1}{L} \\ 0
    \end{bmatrix},
    \quad
    C_a = \begin{bmatrix}
        1 & 0 & 0 & 0
    \end{bmatrix}.
\]

The control input \( u \) for the SFCIA system is defined as follows.
\begin{equation}
    u = -K_c x - K_i w,
\end{equation}
or equivalently:
\begin{equation}
    u = -K_a x_a,
\end{equation}
where \( K_a \) is the combined gain matrix incorporating both state feedback and integral gains.

The system ensures all state variables, including position (\( \theta \)), speed (\( \dot{\theta} \)), and current (\( i \)), are directly measurable using standard sensors. The integral state (\( w \)) is derived from the error \( e = r - y \), allowing straightforward implementation without requiring an observer. This configuration guarantees accurate steady-state performance and system stability.

The block diagram of the State Feedback Controller with Integral Action is presented in Fig.~\ref{fig:SFCIA_bd}, showing how integral action eliminates the steady-state error.

\section{Results and Discussion}

\begin{figure*}[!t]
    \centering
    \begin{minipage}[b]{0.48\textwidth}
        \centering
        \includegraphics[width=\textwidth, height=0.3\textheight, keepaspectratio]{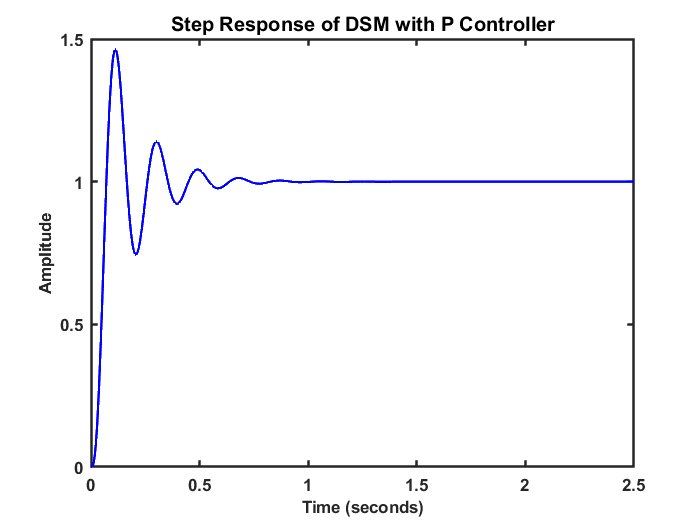}
        \caption{Position Control Step Response of the DSM with P Controller.}
        \label{fig:ContP}
    \end{minipage}
    \hspace{0.02\textwidth}
    \begin{minipage}[b]{0.48\textwidth}
        \centering
        \includegraphics[width=\textwidth, height=0.3\textheight, keepaspectratio]{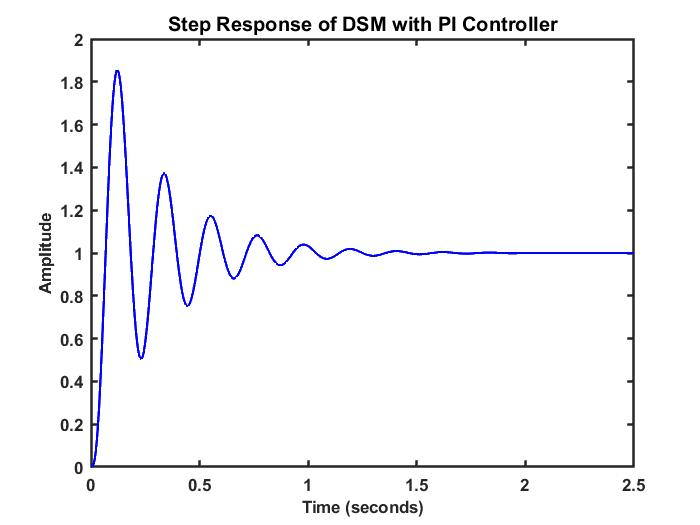}
        \caption{Position Control Step Response of the DSM with PI Controller.}
        \label{fig:ContPI}
    \end{minipage}

    \vspace{0.02\textheight}

    \begin{minipage}[b]{0.48\textwidth}
        \centering
        \includegraphics[width=\textwidth, height=0.3\textheight, keepaspectratio]{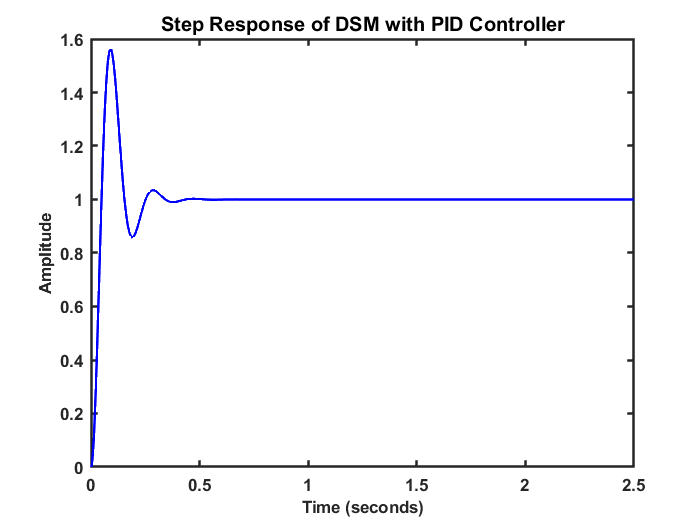}
        \caption{Position Control Step Response of the DSM with PID Controller.}
        \label{fig:ContPID}
    \end{minipage}
    \hspace{0.02\textwidth}
    \begin{minipage}[b]{0.48\textwidth}
        \centering
        \includegraphics[width=\textwidth, height=0.3\textheight, keepaspectratio]{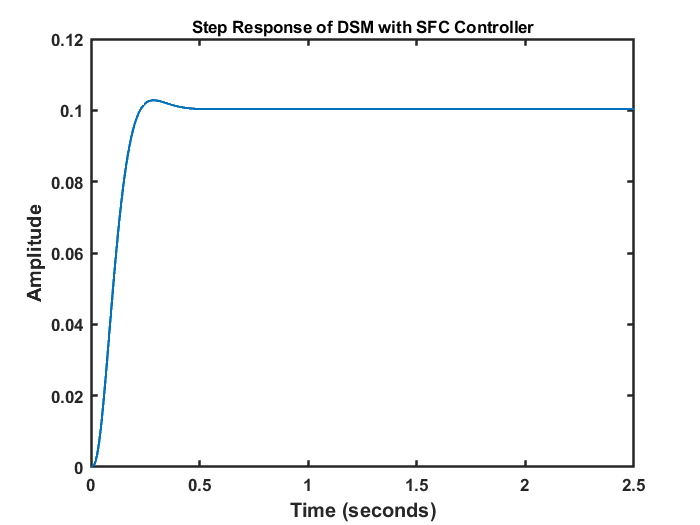}
        \caption{Position Control Step Response of the DSM with SFC Controller.}
        \label{fig:ContSFC}
    \end{minipage}

    \vspace{0.02\textheight}

    \begin{minipage}[b]{0.48\textwidth}
        \centering
        \includegraphics[width=\textwidth, height=0.3\textheight, keepaspectratio]{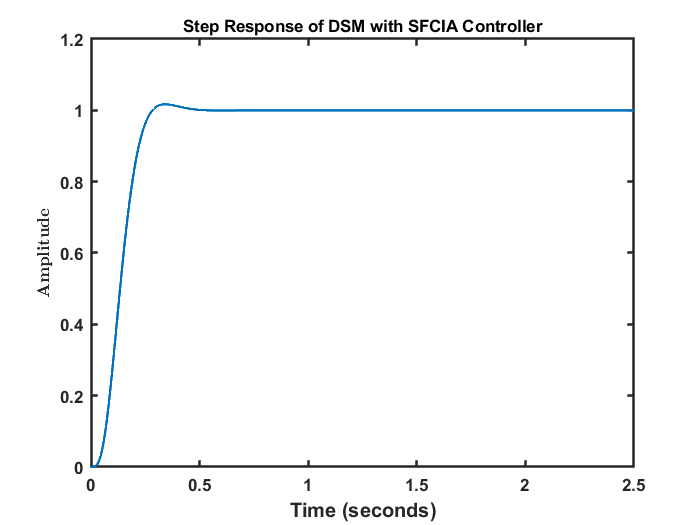}
        \caption{Position Control Step Response of the DSM with SFCIA Controller.}
        \label{fig:ContSFCIA}
    \end{minipage}
    \hspace{0.02\textwidth}
    \begin{minipage}[b]{0.48\textwidth}
        \centering
        \includegraphics[width=\textwidth, height=0.3\textheight, keepaspectratio]{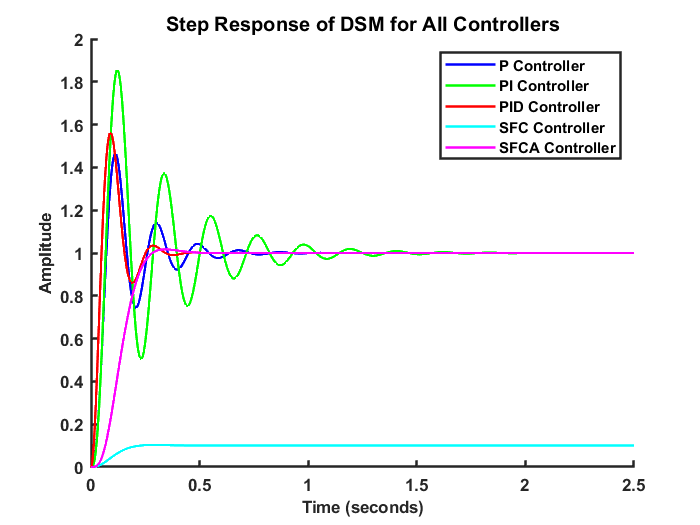}
        \caption{Comparison of the Position Control Step Response of the DSM between all of the controllers.}
        \label{fig:ContCombining}
    \end{minipage}

    \label{fig:controllers}
\end{figure*}

The effectiveness of five controllers (P, PI, PID, SFC, and SFCIA) in position control is assessed using key performance metrics, including maximum overshoot, steady-state error, settling time, rise time, and peak time. The results are summarized in Table~\ref{tab:position}, and the corresponding step response plots are illustrated in Figs.~\ref{fig:ContP}--\ref{fig:ContCombining}.

\begin{table}[h!]
\centering
\caption{Comparison of Controllers for Position Control.}
\label{tab:position}
\begin{tabular}{|l|c|c|c|c|c|}
\hline
\textbf{Metric}              & \textbf{P}   & \textbf{PI}   & \textbf{PID}  & \textbf{SFC}  & \textbf{SFCIA} \\ \hline
Maximum overshoot (\%)       & 46.31        & 85.41         & 56.13         & 2.39          & 0.00           \\ \hline
Steady-state error           & 0.00         & 0.00          & 0.00          & 0.90          & 0.00           \\ \hline
Settling time (sec)          & 0.60         & 1.11          & 0.32          & 0.33          & 0.26           \\ \hline
Rise time (sec)              & 0.04         & 0.04          & 0.03          & 0.13          & 0.15           \\ \hline
Peak time (sec)              & 0.11         & 0.12          & 0.09          & 0.29          & 0.34           \\ \hline
\end{tabular}
\end{table}

The Proportional (P) controller reaches a settling time of 0.60 seconds and successfully eliminates the steady-state error. However, as shown in Fig.~\ref{fig:ContP}, it exhibits a significant overshoot of 46.31\%, along with a rise time of 0.04 seconds and peak time of 0.11 seconds.These traits lead to oscillations, which makes the P controller less ideal for applications that require high precision.

The Proportional-Integral (PI) controller eliminates steady-state error but suffers from the highest overshoot of 85.41\% and the longest settling time of 1.11 seconds, as depicted in Fig.~\ref{fig:ContPI}. Although the rise time is 0.04 seconds and the peak time is 0.12 seconds, excessive overshoot limits its practical use for speed and stability.

The Proportional-Integral-Derivative (PID) controller delivers a well-balanced solution, effectively managing both response time and overshoot in control system applications. As illustrated in Fig.~\ref{fig:ContPID}, it achieves a moderate overshoot of 56.13\% and the shortest settling time of 0.32 seconds. With a rise time of 0.03 seconds and a peak time of 0.09 seconds, the controller shows a quick response, making it well-suited for applications that demand speed, accuracy, and stability.

The State Feedback Controller (SFC) demonstrates improved performance parameters with a low overshoot of 2.39\% and a settling time of 0.33 seconds, as shown in Fig.~\ref{fig:ContSFC}. It introduces a steady-state error of 0.90, limiting precision. With a rise time of 0.13 seconds and a peak time of 0.29 seconds, SFC has a slower initial response than the PID controller, making it better suited for stability-focused systems that can tolerate small steady-state errors.

The State Feedback Controller with Integral Action (SFCIA) achieves the best overall performance. As seen in Fig.~\ref{fig:ContSFCIA}, it attains zero overshoot, eliminates steady-state error, and has the quickest settling time of 0.26 seconds. Although the rise time of 0.15 seconds and the peak time of 0.34 seconds are marginally slower than those of the PID controller, the system excels with zero overshoot, zero steady-state error, and the fastest settling time of 0.26 seconds, ensuring unmatched precision and stability over time. As a result, the SFCIA is ideal for applications demanding high levels of control accuracy.

A combined comparison of all controllers is shown in Fig.~\ref{fig:ContCombining}. The SFCIA controller clearly stands out among the others by demonstrating the highest level of precision in position control. The PID controller provides a well-considered balance among speed, overshoot, and settling time, rendering it an adaptable option for a wide array of general-purpose systems. In contrast, the P and PI controllers exhibit high overshoot and longer settling times, limiting their effectiveness for stability-critical tasks. The SFC controller delivers consistent performance with little overshoot but may result in minor steady-state errors.

In conclusion, the SFCIA controller is recognized as the most effective solution for achieving precise position regulation. It offers zero overshoot, zero steady-state error, and a rapid settling time, which collectively render it an ideal choice for systems where precision is critical. The Proportional-Integral-Derivative (PID) controller is a viable option for applications where a quick response time and moderate overshoot are permissible. Conversely, the Sliding Mode Control (SFC) controller is more appropriate for systems that emphasize stability at the expense of precision.

\section{Conclusion}

This research provided an in-depth evaluation of different control strategies for DC Servo Motors (DSM) by constructing a mathematical model and examining their performance. The controllers investigated include Proportional (P), Proportional-Integral (PI), Proportional-Integral-Derivative (PID), State Feedback Control (SFC), and State Feedback Control with Integral Action (SFCIA). The Ziegler-Nichols method was used to tune the classical controllers, while state-feedback controllers were designed through pole-placement techniques. MATLAB simulations were performed to evaluate these controllers using important performance metrics, including overshoot, steady-state error, settling time, rise time, and peak time.

The findings demonstrate that the SFCIA controller provided the highest level of performance, with no overshoot, no steady-state error, and the quickest settling time. These characteristics make SFCIA particularly suitable for applications demanding high precision, stability, and dynamic performance. The PID controller exhibited a balanced response, achieving zero steady-state error and a fairly quick settling time, which makes it a suitable choice for a variety of general applications. In contrast, the P and PI controllers exhibited significant overshoot and longer settling times, limiting their utility in systems requiring strict precision. The SFC controller achieved low overshoot and fast response but introduced a small steady-state error, making it suitable for tasks where stability is prioritized over absolute precision.

For future endeavors, incorporating advanced optimization techniques, such as machine learning, can significantly enhance controller performance through automated tuning capabilities. Conducting experimental validation on physical DSM systems would assist in tackling real-world issues such as noise, disturbances, and non-linear behaviors within the system. Additionally, exploring adaptive and robust control approaches can further improve performance in dynamic operating conditions. Using energy efficiency as a performance metric will enhance evaluation, especially for energy-sensitive applications.

This study establishes a systematic framework for analyzing and comparing control strategies for DSMs. By combining theoretical analysis with simulation results, it provides valuable insights into precision position control and lays the groundwork for further advancements in industrial and automation systems.

\end{document}